\documentstyle[preprint,aps,eqsecnum]{revtex}

\begin{document}
\input epsf.sty

\def\bfix{~\newline\centerline{XXXXXXXXXXX corrected versionXXXXXXXXXXXXXX}\newline}
\def\efix{~\newline\centerline{XXXXXXXXXXXX corrected version ends hereXXXX}\newline}
\def\bXX{~\newline \centerline{XXXXXXXXXXXXXX this part is correctedXXXXXXX}\newline}
\def\eXX{~\newline \centerline{XXXXXXXXXXXXXXXXXXXXXXXXXXXXXXXXXXXXX}\newline}
\def\ket{{\rangle}}
\def\bra{{\langle}}
\def\ie{{\it i.e.}}
\def\be{\begin{equation}}
\def\ee{\end{equation}}
\def\ba{\begin{eqnarray}}
\def\ea{\end{eqnarray}}
\def\mref#1{Eq.(\ref{Eq:#1})}
\def\mreff#1{Fig.\ref{Eq:#1}}
\def\mreft#1{Table\ref{Eq:#1}}
\def\mlab#1{\label{Eq:#1}}
\def\mlabf#1{\label{Eq:#1}}
\def\mlabt#1{\label{Eq:#1}}
\def\half{\frac{1}{2}}
\def\to{\rightarrow}
\def\nn{\nonumber\\}
\def\sk{\vskip 1cm}
\def\skk{\vskip 3mm}
\def\mat#1#2#3{\langle{#1}\vert{#2}\vert{#3}\rangle}
\def\etal{{\it et al.}}
\def\etc{{\it etc.~}}
\def\eg{{\it e.g.~}}

\def\ibid#1#2#3{{\it ibid.}{\bf #1} #2 {(#3)} }
\def\PR#1#2#3 {{\it Phys. Rev. }{\bf D#1} #2 {(#3)} }
\def\PRL#1#2#3 {{\it Phys. Rev. Lett. }{\bf #1} #2 {(#3)} }
\def\PL#1#2#3 {{\it Phys. Lett. }{\bf #1} #2 {(#3)}  }
\def\AP#1#2#3 {{\it Ann, Phys. }{\bf #1} #2 {(#3)} }
\def\ZP#1#2#3 {{\it Z. Phys. }{\bf #1} #2 {(#3)} }
\def\NP#1#2#3 {{\it Nucl. Phys. }{\bf #1} #2 {(#3)}  }
\def\MPL#1#2#3 {{\it Mod. Phys. Lett.}{\bf #1} #2 {(#3)}  }
\def\NC#1#2#3 {{\it Nuov. Cimm. }{\bf #1} #2 {(#3)}  }
\def\PREP#1#2#3 {{\it Phys. Report }{\bf #1} #2 {(#3)}  }
\def\PROG#1#2#3 {{\it Prog. Theor. Phys. }{\bf #1} #2 {(#3)}   }
\def\SOV#1#2#3{{\it Sov. J. Nucl. Phys. }{\bf #1} #2 {(#3)}   }
\def\JETP#1#2#3{{\it JETP}{\bf #1} #2 {(#3)}   }
\def\RMP#1#2#3{{\it Rev. Mod. Phys.}{\bf #1} #2 {(#3)}   }

\def\phiout{{\phi_a^{out}}}
\def\phiin{\phi_a^{in}}
\def\psiout{\psi_a^{out}}
\def\psiin{\psi_a^{in}}
\def\vin{\phi_{a\mu}^{in}}
\def\vout{\phi_{a\mu}^{out}}
\def\ofx{{(x)}}
\def\op{{\bf P}}
\def\oc{{\bf C}}
\def\ot{{\bf T}}
\def\cp{{\bf CP}}
\def\cpt{{\bf CPT}}
\def\vecr{{\vec r}}
\def\vecn{{\vec {\nabla}}}
\def\vecA{{\vec A}}
\def\psibar{\overline\psi}
\def\outin{{{out}\choose{in}}}
\def\inout{{{in}\choose{out}}}

\def\vr{\vec x}
\def\itt{{\it T}}
\def\b{{\bf b}}
\def\a{{\bf a}}
\def\d{{\bf d}}

\def\alphadot{{\dot\alpha}}
\def\betadot{{\dot\beta}}
\def\gammadot{{\dot\gamma}}

\def\N{\sqrt{{{E+mc^2}\over{2mc^2}}}}
\def\F#1{{{#1}\over{E+mc^2}}}
\def\itemm{\hangindent\parindent\textindent}
\def\noi{\noindent}
\def\onehead#1{\vskip1pc\leftline{\bf #1}}
\def\twohead#1{\vskip1pc\leftline{\bf #1}}
\def\ts{\thinspace}

\def\sq2{{1\over{\sqrt{2}}}}
\def\omegaar{{\vec{\omega}}}
\def\kbar{\overline K}
\def\Pbar{\overline P}
\def\Abar{\overline A}
\def\dbar{\overline d}
\def\ubar{\overline u}
\def\sbar{\overline s}
\def\bbar{\overline B}
\def\gbar{\overline g}
\def\pbar{\overline P}
\def\qbar{\overline q}
\def\vecr{\vec r}

\def\dm{\Delta m}
\def\ss{(1+s^2)}
\def\cdmp{cos\Delta m(t_1+t_2)}
\def\cdm{cos\Delta m(t_1-t_2)}

\def\g5{\gamma_5}
\def\gm{(1-\gamma_5)}
\def\gp{(1+\gamma_5)}

\def\mvec#1{\vec{#1}\,}
\def\pslash{\mbox{/\llap p}}
\def\slash#1{\mbox{/\llap #1}}

\def\msmall#1{\mbox{\rm \small #1}}
\newcommand{\matel}[3]{\langle #1|#2|#3\rangle}
\newcommand{\hscale}{\mu\ind{hadr}}
\newcommand{\aver}[1]{\langle #1\rangle} 
\renewcommand{\Im}{\mbox{Im}\,}
\renewcommand{\Re}{\mbox{Re}\,}
\newcommand{\GeV}{\,\mbox{GeV}}
\newcommand{\MeV}{\,\mbox{MeV}}
\newcommand{\BR}{\,\mbox{BR}}

\begin{titlepage}
\renewcommand{\thefootnote}{\fnsymbol{footnote}}

\begin{flushright}
DPNU-99-09\\
\end{flushright}
\vspace{.3cm}
\begin{center} \Large  
{\bf Comments on \cpt~tests\footnote{Talk presented at KEK-Tanashi.}
}
\vspace*{.3cm}
\\
{\Large $ A. I. Sanda $}\\
\vspace{.4cm}
{\normalsize 
{\it  Physics Dept., Nagoya University, 
Nagoya 464-01,Japan}\\

\vspace{.3cm}
e-mail address:  
sanda@eken.phys.nagoya-u.ac.jp }

\vspace*{.4cm}
{\Large{\bf Abstract}}\\
\end{center} 
What do we know about \cpt~ symmetry? We conclude that the direct measurement of \cpt~ violation leads to a result that \cpt~ violating forces is limited 
to about 30\% of \cp~vioating forces. The best test of \cpt~symmetry is to see 
if theoretical prediction for the phases of $\eta_{+-}$ and $\eta_{00}$ agees with experiments.
We discuss uncertainties associated with this prediction, and how to improve it. A new test of \cpt~ is also discussed.
\end{titlepage}

\section{Introduction}

The 'classical' tests of \cpt~symmetry concern the equality of masses 
and lifetimes for particles and antiparticles. The 
most dramatical way of stating the validity of \cpt~symmetry is to
state the 
experimental upper bound\cite{PDG}: 
\be 
\frac{|M_{\overline K^0} - M_{K^0}|}{M_K} < 10^{-18} \; \;.  
\mlab{CPTKKBAR} 
\ee 
This is very misleading. The upperlimit on the right hand side depends on what we choose for the denominator. Why is $M_K$ a reasonable choice? Bulk of K meson mass comes from strong interaction. So, if we compare the strength of \cpt~
violating force with that of QCD, indeed, \cpt~ violating force is very small.
Since \cp~symmetry forces $M_{\overline K^0}= M_{K^0}$, \cpt~symmetry must be tested in the environment where \cp~symmetry is broken.
Then it is perhaps more reasonable state:
\be 
\frac{|M_{\overline K^0} - M_{K^0}|}{\eta\Gamma_{K_S}} < 0.04 \; \;  
\mlab{CPTKKBAR2} 
\ee 
Suddenly, it is no longer so impressive.

In Sec. 2, we first discuss a direct test of \cpt~ using recent experimental result from CPLEAR. In Sec. 3, 
we discuss theoretical prediction of $\eta_{+-}$ and $\eta_{00}$ phases - exposing all theoretical subtleties.  In Sec. 4, we suggest a new \cpt~test which requires only studying $2\pi$ and $3\pi$ decays of $K_L$.

\section{Direct test of \cpt}
The most direct test of \cpt~is to study time dependence of $K_{l3}$ decays.
\ba 
A_{T}(t) &=& \frac{\Gamma (\overline K^0(t) \to e^+) - 
\Gamma ( K^0(t) \to  e^-)}
{\Gamma (\overline K^0 \to e^+) +  
\Gamma ( K^0 \to  e^-)}\nn
A_{CPT}(t) &=& \frac{\Gamma (\overline K^0(t) \to e^-) - 
\Gamma ( K^0(t) \to  e^+)}
{\Gamma (\overline K^0 \to e^-) +  
\Gamma ( K^0 \to  e^+)}\nn
\ea
For $ t\gg \Gamma _S^{-1} $ we have
\be 
A_T(t) + A_{CPT}(t) \simeq 2({\rm Im}\, \phi + 
{\rm Re \, cos}\, \theta ) ,
\mlab{rct}
\ee  
where $\cos\theta\ne 0$ and $\phi\ne 0$ imply \cpt, and \cp~symmetries are violated, respectively.
Since $\epsilon\sim \frac{i}{2}\phi$ we can extract $\Re \cos\theta$.
The most accrate measurement of $\Re\cos\theta$ is give by CPLEAR collaboration
\cite{CPLEAR444}:
\be
\Re\cos\theta=(6.0\pm 6.6_{stat.}\pm 1.2_{syst.})\times 10^{-4}.
\ee
Comparing this number to $\epsilon\sim 2\times 10^{-3}$, we see that \cpt~violating forces is limited to 30\% of \cp~violating forces. While the new experimental result represents a major step toward understanding \cpt~symmetry, this is hardly illuminating.

\section{\cpt~Tests and Phases of $\eta _{+-}$ and 
$\eta _{00}$}
%

In this section, we show that the most stringent and legitimate
 test of \cpt~is provided by the phases of $\eta_{+-}$ and
$\eta_{00}$. It is well known that with \cpt~symmetry, 
 $\phi_{+-}=\phi_{00}=\phi_{SW}$, 
where $\phi_{SW}=\tan^{-1}\left(\frac{2\Delta M}{\Delta\Gamma}\right)$,
to a very good approximation. We want to compute the correction to this relation. To first order in
$\phi _{+-}-\phi_{SW}$ and
$\phi _{00} -\phi_{SW}$,
it is possible to derive the following relation:
\ba 
\frac{|\eta _{+-}|}{ {\rm sin}\phi _{SW}}\left( \frac{2}{3} \phi _{+-} + \frac{1}{3}\phi _{00} 
- \phi _{SW} \right) &=&
-ie^{-i\phi_{SW}}\frac{\sum_{f\not =(2\pi)_0}\epsilon(f)}{\sin\phi_{SW}}
-\half\Re \Delta_0\nn
\epsilon(f)&=&ie^{i\phi_{SW}}\frac{\Im\Gamma_{12}(f)}{\Delta \Gamma}\cos\phi_{SW},
\ea
where $\Delta_I=1-\frac{\overline A_I}{A_I}$, and $A_I$ and $\overline A_I$ are
$K\to(\pi\pi)_I$ and $\overline K\to(\pi\pi)_I$ amplitudes, respectively.
Here $(\pi\pi)_I$ denotes $\pi\pi$ final state with isospin $I$. Finally,
we emphasize that the sum over $f$ does not include $f=(2\pi)_0$
channel.

Let us first consider $\Re~ \Delta_0$.
Since we assume \cpt~ symmetry, we can write
\be
\frac{\overline A_0}{A_0}=e^{-2i\xi_0}
\ee
Then $\Re~ \Delta_0={\cal O}(2\xi_0^2)$ and is second order in \cp~violating 
parameters. These terms are same order as neglected terms.

Now, we estimate $\epsilon(f)$. First some preliminaries.
For those final states $f$ for which $\cpt |f\rangle = 
|f\rangle$ holds, we can write 
\be 
\Gamma _{12}^f \equiv 2\pi \rho _f 
\matel{f}{H_W}{K^0}^2,
\ee
\ba 
\small
{\rm Im}\, \Gamma _{12}^f &=& -i\pi \rho _f  
\left( A_f^2 - A_f^{*2}\right) =\nn 
&=& -i\pi \rho _f 
\left( A_f^2 - \overline A_f^{2}\right) 
\nonumber 
\ea 
where $A_f=\matel{f}{H_W}{K^0}$ and $\overline A_f=
\matel{f}{H_W}{\overline K^0}$,
and we have used the fact that \cpt~ is an anti-linear operator. 
To first order in \cp~violtion,
\be
\epsilon(f)=e^{i\phi_{SW}}\cos\phi_{SW}\frac{\Gamma(K\to f)}{\Delta\Gamma}
\left( 1-\cp_f\frac{\overline A_f}{A_f}\right)
\mlab{iden}
\ee
Now we consider various intermediate states $f$.
\begin{itemize}

\item  $f=(2\pi)_2$  

Here $\epsilon((2\pi)_2)$ can be estimated by noting that
\be
\epsilon'=\frac{1}{2\sqrt{2}}\omega\ e^{i(\delta_2-\delta_0)}(\Delta_0-\Delta_2)\ee
where $\Delta_I=1-\frac{\overline A_I}{A_I}$.
Baring unexpected cancellation between $\Delta_0$ and $\Delta_2$, we estimate that 
\be
|\epsilon((2\pi)_2)|=2\omega\epsilon'\sim 10^{-7}.
\ee
\item $f=(\pi^+\pi^-\pi^0)_{\cp=+}$
\ba
|\epsilon((\pi^+\pi^-\pi^0)_{\cp=+})|&=&\frac{1}{2\sqrt{2}}
Br(K_S\to (3\pi)_{\cp+})
\left|1-\frac{\overline A((3\pi)_{\cp+})}{A((3\pi)_{\cp+})}\right|\nn
&\le&\left(1.2 {{+0.4}\atop {-0.3}}\right)\times 10^{-7}
\ea
where we have used 
$Br(K_S\to\pi^+\pi^-\pi^0)=\left(3.4 {{+1.1}\atop {-0.9}}\right)\times 10^{-7}$
\cite{PDG}.
\item $f=(\pi^+\pi^-\pi^0)_{\cp=-}$  
\be
|\epsilon((\pi^+\pi^-\pi^0)_{\cp=-})=\frac{1}{2\sqrt{2}}Br(K_L\to (3\pi)_{\cp-})\frac{\Gamma_L}{\Gamma_S}
\left(1+\frac{\overline A((3\pi)_{\cp-})}{A((3\pi)_{\cp-})}\right)\nn
\ee
\cp~odd $\pi^+\pi^-\pi^0$ with symmetrized $\pi^+\pi^-$ state is given by
\cite{CPLEAR407}:
\ba 
{\rm Re}\, \eta _{+-0} &=& (-2 \pm 7~{+4\atop -1})\cdot 10^{-3}  
\mlab{Re482}\\
{\rm Im}\, \eta _{+-0} &=& (-9 \pm 9~{+2\atop -1})\cdot 10^{-3} .
\mlab{Im410} 
\ea 
where the first and the second errors are statistical error, and systematic
error, respectively. Using 
$\eta_{+-0}=\half\left(1+\frac{q_1}{p_1}\frac{\overline A(\pi^+\pi^-\pi^0)_{\cp-}}{A(\pi^+\pi^-\pi^0)_{\cp-}}\right)$, we are lead to a reasonable guess:
\be
\left(1+
\frac{\overline A(\pi^+\pi^-\pi^0)_{\cp-}}
{A(\pi^+\pi^-\pi^0)_{\cp-}}\right)\sim \left(1+\frac{q_1}{p_1}
\frac{\overline A(\pi^+\pi^-\pi^0)_{\cp-}}
{A(\pi^+\pi^-\pi^0)_{\cp-}}\right)<2\times 10^{-2}
\ee
which gives
\be
\epsilon(\pi^+\pi^-\pi^0)_{\cp-}\le 1.4\times 10^{-6}.
\ee
\item $f=3\pi^0$

This can be estimated by noting that
\be
\eta_{000}=\half\left(1+\frac{q_1}{p_1}\frac{\overline A(3\pi^0)}{A(3\pi^0)}\right)
\ee
and the experimental value for $\eta_{000}$ is given by\cite{CPLEAR425}:
\ba 
{\rm Re}\, \eta _{000} &=& 0.18\pm 0.14_{stat.} \pm 0.06_{syst.}  
\mlab{Re482}\\
{\rm Im}\, \eta _{000} &=& -0.05\pm 0.12_{stat.} \pm 0.05_{syst.}  
\mlab{Im410} 
\ea 
which leads to
\be
\left|1+\frac{\overline A(3\pi^0)}{A(3\pi^0)}\right|\sim\left|1+\frac{q_1}{p_1}
\frac{\overline A(3\pi^0)}{A(3\pi^0)}\right|<2|\eta_{000}|<.46
\ee
Using \mref{iden}, we obtain:
\be
|\epsilon(3\pi^0)|=
\frac{\Gamma(K_L\to3\pi^0)}{2\sqrt{2}\Gamma_S}\left(1+\frac{\overline A(3\pi^0)}{A(3\pi^0)}\right)<6\times 10^{-5}
\ee

\item
In considering $f = \pi l \nu$ we have to allow for a violation of 
the $\Delta Q = \Delta S$ rule as expressed by the complex parameter 
\be
x=\frac{\bra l^+\nu\pi^-|{\cal H_W}|\overline K\ket}{\bra l^+\nu\pi^-|{\cal H_W}|K\ket}.
\ee
Since
\be 
{\rm Im}\, \Gamma _{12}^{\pi l \nu} \simeq  {\rm Im}\, x 
\Gamma (K\to \pi \mu \nu ), 
\ee 
we find 
\be 
|\epsilon (\pi l \nu )| \leq 4 \cdot 10^{-7}, 
\ee 
where we have used the bound ${\rm Im}\, x = (0.5 \pm 2.5)\times 10^{-3}$ 
from \cite{CPLEAR444} 
and $Br(K_S\to\pi^\mp\mu^\pm\nu)\sim 5\times 10^{-4}$. 

\end{itemize}
In summary: we can conclude that the upperlimit is dominated by $3\pi^0$ 
channel, and we have
\be 
| \sum _f \epsilon ^f | \leq \sum _f |\epsilon ^f| \leq 
 6 \times 10^{-5}. 
\ee 
With our new evaluation of theoretical error we obtain\cite{carosi,adler6}:
\be 
\frac{2}{3}\phi _{+-} + \frac{1}{3}\phi _{00} - \phi _{SW} 
\simeq (0.17 \pm 0.81_{exp}+1.7_{theory})^o 
\ee 
So, we have about $\frac{1.7^o}{\phi_{SW}}\sim 4\%$ test of \cpt~violation.


\section{New test of \cpt~ symmetry}
Define the rate asymmetry
\be
{\cal A}(t)=\frac{\Gamma(K(t)\to\pi^+\pi^-\pi^0)-
\Gamma(\overline K(t)\to\pi^+\pi^-\pi^0)}{\Gamma(K(t)\to\pi^+\pi^-\pi^0)+
\Gamma(\overline K(t)\to\pi^+\pi^-\pi^0)}
\ee
Noting  that, in general,
\ba
|K^0\ket&=&\frac{1}{p_1q_2+q_2p_1}[q_1|K_L\ket+q_2|K_S\ket]\nn
|\overline K^0\ket&=&\frac{1}{p_1q_2+q_2p_1}[-p_1|K_L\ket+p_2|K_S\ket]\nn
\ea
we see that decay $K\to\pi^+\pi^-\pi^0$ at $t\to\infty$
leads to an asymmetry 
\be
\lim_{t\to\infty}{\cal A}(t)
=1-\left|\frac{q_1}{p_1}\right|^2.
\ee
This is a curious result. If we look at $K\to\pi^+\pi^-$ at $t\to\infty$,
which is $K_L\to\pi^+\pi^-$, we obtain
\be
\epsilon\sim\left(1-\frac{q_2}{p_2}\right).
\ee
So, by comparinhg  $t\to\infty$ limit of $K\to\pi^+\pi^-$ and 
$K\to\pi^+\pi^-\pi^0$ decays we have a test of \cpt~symmetry:
\be
\frac{q_2}{p_2}\stackrel{?}{=}\frac{q_1}{p_1}
\ee
\section{Summary}
We have examined existing tests of \cpt~symmetry. We found that strength of \cpt~violating interaction is bounder by 4\% of the \cp~violating interaction.
Thed majority of the theoretical error comes from the uncertainty in 
$\eta_{000}$. We recomend some effort in improving this measurement.
A new test of \cpt~ comparing \cp violation in $K_L\to 3\pi$ and $K_L\to 2\pi$
is suggested.

\sk
\centerline{Acknowledgements}
This work has been supported in part by Grant-in-Aid for Special Project Research 
(Physics of \cp~violation).

\end{document}